# Superconductivity in Quantum Complex Matter: the Superstripes Landscape

Antonio Bianconi[1,2,3]

[1] Rome International Center for Materials Science Superstripes RICMASS, via dei Sabelli 119A, 00185 Roma, Italy
[2] Institute of Crystallography, CNR, via Salaria, Monterotondo, Roma, I-00015, Italy
[3] National Research Nuclear University MEPhI (Moscow Engineering Physics Institute), 115409 Moscow, Russia

Abstract

While in XX century the theory of superconductivity has focused on a homogeneous metal with a rigid lattice which can be reduced to a single effective conduction band in the dirty limit. Today in the XXI century, the physics of superconductivity is focusing on complexity of quantum matter where novel quantum functionalities with lattice inhomogeneity at nanoscale (between 1 nm and 100 nm) and at mesoscopic scale (in the range 100-10000 nm), where the electronic structure need to be described by multiple Fermi surfaces and multiple gaps in the superconducting phase in the clean limit. The present issue of Journal of superconductivity and Novel magnetism collects papers presented at the International Conference Superstripes 2019 which was held on June 23-29, 2019, in Ischia Island, Italy. The series of Stripes conference started on Dec 8 1996 and it has contributed to scientific advances in the new physics of superconductivity in quantum complex systems in these last 23 years where polarons, strain, multigap superconductivity, exchange interaction between condensates, granular superconductivity and percolation play a key role. The articles collected in this issue cover hot topics of the new quantum physics of complex matter including the mechanism of high temperature superconductivity, complex magnetic orders and orbital physics in strongly correlated materials which have been growing rapidly in these last two years opening new perspectives also in mesoscopic quantum engineering.

## 1. Introduction

Alex Müller[1,2] discovered in 1986 high temperature superconductivity (HTS) in doped perovskites searching HTS in unconventional metals in the strong coupling approximation beyond BCS theory. In the BCS theory the condensate is made of Cooper pairs, i.e., two high energy fermions interacting by exchange of a low energy boson with zero mass. On the contrary in the Alex proposal the HTS condensate is formed by Jahn-Teller bipolarons[3] more close to Bose-Einstein Condensate (BEC) than to the Bardeen–Cooper–Schrieffer (BCS) condensate[4]. The key idea of Alex which led him to the discovery was that a novel non-BCS superconducting phase could appear near the insulator to metal transition (MIT) in perovskite materials where quantum mechanics of many body systems could take advantage of the intricate fluctuating lattice inhomogeneity[2,5] i.e., the correlated dynamical local lattice fluctuations. Today the physics of nanoscale and mesoscale heterogeneity of perovskites is of high interest not only for understanding superconductivity in quantum complex matter but also in the neighbor fields of mesoscopic quantum physics[6] and quantum engineering[7].

After the 1987 March meeting in New York the theoretical physics community focusing its interest on the fundamental mechanism of high temperature superconductivity has been dominated by the popular paradigm followed by the majority of scientists: HTS superconductivity occurs in correlated electronic systems in a rigid periodic homogeneous lattice described by the single band Hubbard model with charge and spin interaction but negligible role of electron-phonon interaction.

For many years the unexpected experimental features found by experimental physics which falsify this paradigm have been assigned by the majority to *intrinsic effects due to* disorder induced by competing interactions and have been called spin-charge stripes, intertwined orders, nematicity[8].

On the contrary experimental results have shown the emerging of a new physics involving lattice effects, structural phase transitions, local lattice distortions, anharmonic modes, multiple orbitals and spin-orbit interactions which have been called *extrinsic effects* by the majority. However experimental evidence for the key ubiquitous essential extrinsic effects was accumulated over 33 years: lattice fluctuations near structural phase transitions, phase separation, bond fluctuations, local lattice distortions, inhomogeneity, defects distribution and their self organization, uniaxial and isotropic pressure effects, lattice misfit, micro-strain, electron-phonon interaction, anharmonicity, polarons, stripes of distorted lattice, granular superconductivity, percolation phenomena, proximity of the chemical potential to Lifshitz transitions.

A group of few developed alternative theories where the so called *extrinsic features* are the base of the fundamental mechanism of high temperature superconductivity. This growing community has falsified all assumptions of the popular paradigm of the majority based on a homogeneous system made of a single effective electronic component with strong correlation and only charge-spin interaction.

The few have shown that multi-band Hubbard models are needed to grab key details of the physics of cuprates, organics and iron based superconductors. Multi orbital models are needed for all high temperature superconductors including cuprates, diborides, doped fullerene and pressurized hydrides, Quantum configuration interaction between non degenerate orbitals giving pseudo-Jahn Teller polarons, local anharmonic modes and vibronic coupling is of high relevance.

The minority group followed different roadmaps in contrast with the paradigm of the majority "Superconductivity in a single correlated band in a homogeneous lattice". A new common paradigm emerged clearly at the 2004 Stripes conference which was titled "Nanoscale heterogeneity and quantum phenomena in complex matter". In 2020 the scientific community which was for 30 years a minority is becoming the majority with a new paradigm "Superconductivity in Quantum Complex Matter" (SQCM).

The few who started the investigation of SQCM included Nobel prize winners like J. Bardeen[9], T. D. Lee[10], A. J. Heeger[11], V. L. Ginzburg[12-15], J. B., Goodenough[16-19], and P. G de Gennes[20] and many outstanding scientists like, J. Friedel[21,22], G. Deutscher[23], A.J. Jorgensen[24], A.R. Bishop[25], E. Teller[26], J. Ashkenazi and C.G. Kuper[27], P. Radaelli[28], P.C. Hammel and D. J. Scalapino[29], T. Egami[30-33], J. C Phillips[34] G. Shirane[35], C.P. Slichter[36], V.Z. Kresin[37,38], Y.N. Ovchinnikov[39] N.W. Ashcroft[40,41], L.P. Gor'kov[42,43], F. Iachello[44] to cite few of them.

## 2. The first Stripes conference in 1996

The series of stripes conferences with related workshops in Rome, Erice and Ischia have been addressed to develop scientific debate in this community.

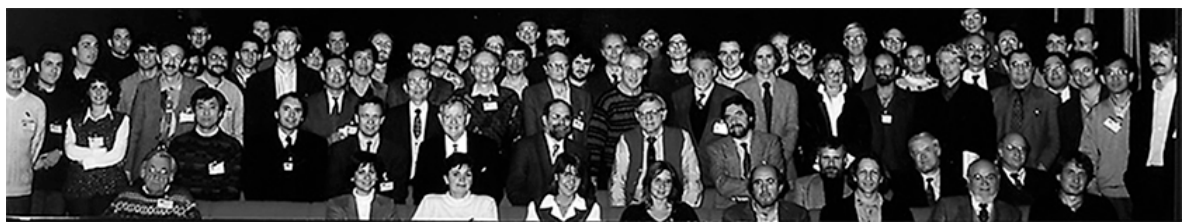

**Fig. 1.** First international conference *Stripes and Lattice Instabilities*, Rome, Dec 8-12, 1996

The first stripes conference (Fig. 1) was held at Rome University in December 8-12, 1996. The experiments, made in the first ten years from the discovery [1], had provided evidence for both "multiple electronic components" and "phase separation" [45-47]. Experimental evidence for extrinsic stripes made of different lattice bond lengths and tilting distribution in the $CuO_2$ plane of $Bi_2Sr_2CaCu_2O_{8+y}$ (Bi2212) where reported in 1992 first at the Erice workshop [46] and in 1993 at the Cottbus workshop [47], The experiments were between the first results obtained by using the novel ESRF synchrotron radiation facility based on joint use of diffraction and the new local and fast probe of instantaneous Cu-O bond distribution called Cu K-edge X-ray Absorption Near Edge Structure (XANES) and Extended X-ray Absorption Fine Structure (EXAFS) [48,49]. A patent of quantum engineering for design of quantum coherent heterostructure at atomic limit (superlattices made of nanoscale units, quantum wells or quantum wires) forming heterogeneous high temperature superconductors [50,51] was submitted with priority date on Dec 7 1993. The nanoscale stripe units induce quantum confinement since their size is of the order of the electron Fermi wavelength and the macroscopic quantum coherence is controlled by their separation smaller than the superconducting coherence length. At the first Stripes conference the earlier experimental results [48,49] giving the mesoscopic stripes structure have been confirmed by high resolution polarized EXAFS data [52-54] and the first resonance elastic X-ray diffraction experiment of cuprates collected at ESRF [55]. The theory group of Perali and Valletta developed in 1995-1996 the multigap Bogoliubov theory for superconducting nanoscale heterostructure with quantum size effects and the gaps in the quantum subbands, critical temperature and isotope effects in the numerical calculations of multigap superconductivity near a Lifshitz transition.

The Bianconi-Perali-Valletta BPV theory presented at the first stripes conference [56-58] provided an innovative theoretical approach for describing 2D superconductivity in a complex mesoscopic landscape made of quantum wires. This was a new proposal for 150K temperature superconductivity showing the superconducting dome with a Fano line shape given by the quantum shape resonance between superconducting gaps near a Lifshitz transition. The results had shown that the positive role of the heterogeneity of Fermi surface topology in the k-space with a first condensate in the BCS-BEC crossover near a Lifshitz transition coexisting with another BCS condensates. The theory included both Cooper pairs formation by phonon-exchange and the Majorana attractive or Heisenberg repulsive exchange interactions calculated by first principle between condensates which were not included in the BCS theory.

## 3. The Stripes conferences in 1998-2004

The second stripes conference was held in 1998 at Aula Magna of Sapienza University in Rome since the size the meeting was increasing with hundreds of participants and many world leaders in the field including Müller, Gor'kov, Shirane, and Emery attracted by the Rome international forum open to confrontation between competing new theories and new experimental results on stripes physics. The Proceedings of the stripes conference have been published in book titled "Stripes and Related Phenomena" [59] (Fig. 2).

At Stripes 1998 Müller presented an important key opening talk titled *From Phase Separation to Stripes* [60] pointing out that the emerging mesoscale stripes landscape was in agreement with previous results showing the presence of pseudo-Jahn Teller (PJT) polarons, and phase-separation in fact first nanoscale stripes of linearly self organized JPT polarons are intercalated by second coexisting stripes of Fermi particles.

Further advances in the BPV theory, presented at Stripes 1998, provided evidence that by tuning the chemical potential near a Lifshitz transition of the type called *opening a neck* in a subband formed by quantum confinement, the predicted dome of the critical temperature tracks the shape resonance with the Fano line-shape profile and reaches a maximum of about 130 Kelvin as the experimental maximum $T_c$ in cuprates [61]. A major new result presented at Stripes 98 conference was the direct visualization of the pseudogap topology in Bi2212 Fermi surface in the normal phase indicated by missing segments of the Fermi surface around (π,0) provided by ARPES data collected in the k-scanning mode [62]. In the communication on Lattice-Charge Stripes in the High-$T_c$ Superconductors [63] it was first proposed a thermodynamic 3D phase diagram for the superconducting atomic $CuO_2$ layers in all cuprate families where the critical temperature is plotted as a function of both doping and chemical pressure.

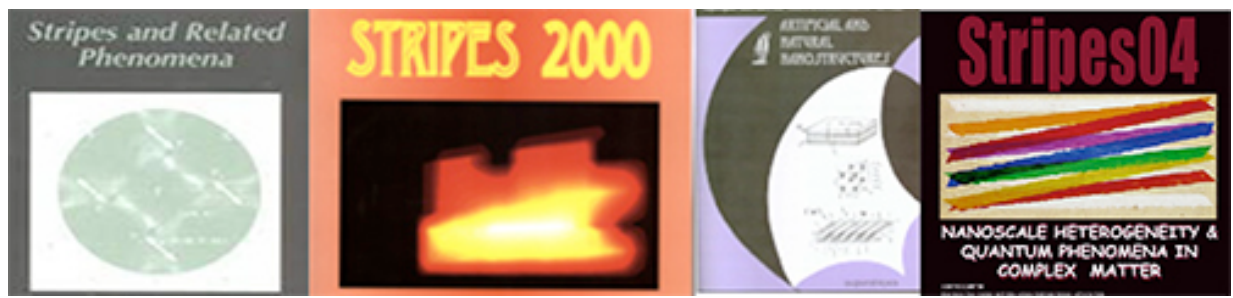

**Fig. 2.** Posters of the Stripes international conferences, from left to right: i) *Stripes and Related Phenomena* 1998; ii) Conference Stripes 2000, iii) Rome conference on Artificial and Natural Heterostructures 2001, iv) Conference Stripes 2004, Nanoscale Heterogeneity and Quantum Phenomena in Complex Matter.

These experimental results support the theoretical prediction by the Bianconi-Perali-Valletta theory of a dome both by changing the charge density as well as the lattice parameters. The BPV theory shows that the maximum of the dome of the superconducting temperature occurs by tuning the chemical potential by charge density and pressure in the proximity of a topological transition of the Fermi surface called Lifshitz transition of the type opening a neck measured in ARPES. The new 3D phase diagram was presented on July 2000 at the in conference on Physics in Local Lattice Distortions held in Ibaraki, Japan, 23-26 July 2000 [64] where it was pointed out the proximity of the maximum $T_c$ with with a Lifshitz transition [65]. The new 3D phase diagram [66] was one of the main results presented at Stripes2000 conference held in Rome on Sept 25-30 2000 [67]. Moreover the analysis of the structural data for materials close to the top of the 3D dome have shown the presence of critical charge, lattice and spin fluctuations near a strain quantum critical point showing self-organization of local lattice distortions forming puddles of superlattice of quantum wires forming a mesoscopic landscape called "superstripes" [68] like in a material showing a critical opalescence that therefore should favor the amplification of the critical temperature [69]. In December 2000 in the same days when these results were published, Nagamatsu, a student in the Akimitsu lab, where the group was studying superconductivity mediated by magnetic interactions, measured 39K superconductivity in a commercial sample of $MgB_2$. The announcement was made in January 2001 and the paper was published in Nature [70] on March 1st 2001 in same day when we submitted two papers showing two gaps in the multigap superconductivity of $MgB_2$ where the maximum Tc occurs at the critical charge density in the atomic boron layers and at the critical strain, due to the critical chemical pressure for all families of diborides [71,72] which was supported by a thermal conductivity experiment [73].

On March 12, 2001 at APS *March-Meeting* in *Seattle*, Washington, over a thousand people packed into the Grand Ballroom of the Westin Hotel at a post-deadline "Session on MgB2". The session began at 8 pm on March 12. Interest was intense, although the crowd had dwindled to perhaps a couple of hundred when the 79th and final paper was presented at about 1:15 am. Only one of the 79 talks, the last 79th talk proposed that $MgB_2$ was not a conventional homogeneous 3D single gap BCS superconductor in the dirty limit, but the exotic realization of multigap anisoptropic superconductivity in the clean limit tuned at a shape resonance. This simply binary alloy is actually a composite material, a heterostructure at atomic limit, i.e., a superlattice of van der Waals atomic layers of boron intracalated by Mg layers like intercalated graphite superconductors. The 39K superconductivity appears only at a critical value of the micro-strain and charge density in the atomic boron layer.

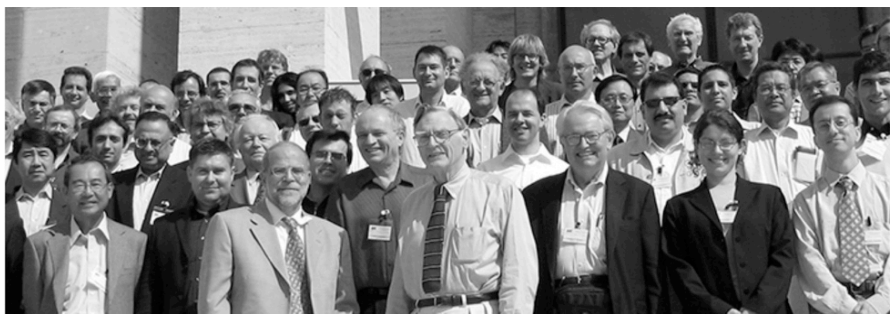

**Fig. 3.** Int. Conference Stripes 2004, Nanoscale Heterogeneity and Quantum Phenomena in Complex Matter, Rome, Italy Sept. 26 - Oct. 2 2004

At this particular point of the 3D phase diagram (temperature, density, pressure) proposed in cuprates tuned at the maximum of the shape resonance in the superconducting boron layers the chemical potential near a Lifshitz transition in the band is tuned at the shape resonance in the superlattice of atomic boron layer where the chemical potential is tuned at the right distance from the Lifshitz transition for appearing of the new sigma Fermi surface (type I) where $T_c$ is very low and, near the neck opening Lifshitz transition in the small sigma Fermi surface. In our lab we prepared many samples with the substitution of Sc and Al for Mg tunimg both the charge density and strain [74,75] which allows to tune the system to test the critical temperature dome with the maximum at 39K where the shape resonance shows a maximum. The fine tuning of the chemical potential with pressure and chemical doping has been confirmed at the Superstripes 2019 conference by Alarco [76].

At the Stripes 2004 conference (Fig. 3) the complex mesoscale landscape called "Superstripes" proposed in 2000 for high temperature superconductors was supported by several experimets an theoretical models. The evidence for the expected nanoscale phase separation and the formation of the mesoscopic superstripes spatial heterogeneous landscape in doped $MgB_2$ was obtained by neutron [77] and x-ray diffraction [78]. In 2008 Hosono discovered high temperature superconductivity in iron based superconductors [79]. These are perovskite materials where atomic Fe superconducting layers are intercalated with fluorite spacer layers like in electron doped cuprates giving large Fe-Fe distances which because of the Hund rule allows the coexistence of multiple orbitals at the Fermi level. It was therefore clear the need of the multiorbital Hubbard models for the iron-based superconductors. [80] Superconductivity in diborides in iron based superconductors has provided evidence for the realization of multigap superconductivity in heterostructures at atomic limit made of metallic atomic layers intercalated by spacer layers described in our patent therefore Perali, Valletta, Innocenti, and Caprara developed the BPV theory for a superlattice of quantuim wells [81,82] predicting the evolution with doping by tuning the chemical potential around type I and type II lifshitz transitions for opening a neck in tubular Fermi surface [81,82].

## 4. The Superstripes 2019 conference

At Superstripes 2019 conference Torsello et al [83], Pal et al. [84] and Ptok et al. [85] presented the most recent investigations on the gap anisotropy in multigap superconductivity and the effect of pressure in iron based superconductors.

Organic superconductors form another class of superconductors showing the realization of a superconducting dome of multigap superconductivity in superlattices of quantum wires described by Mazziotti [86]. At Superstripes 2019 Tsuchiya et al [87] and Nakagawa et al, [88] have reported the most recent results on the investigation of carrier dynamics and phase separation by spectrally resolved pump-probe spectroscopy and polarized time-resolved spectroscopy of electronic phase separation. Arrested nanoscale phase separation was predicted for multiband correlated metals near a topological Lifshitz transition [89] which is reached by tuning strain and charge [90] with a complex landscape of polaronic short range charge density waves [91].

Hideo Aoki [92] has developed the multi band Hubbard model in the case of coexistence near the Fermi level of a flat band with virtual pair-hopping processes in which pairs are scattered between the flat band spots and Cooper pairs in the dispersive band to explain superconductivity in twisted graphene bilayers.

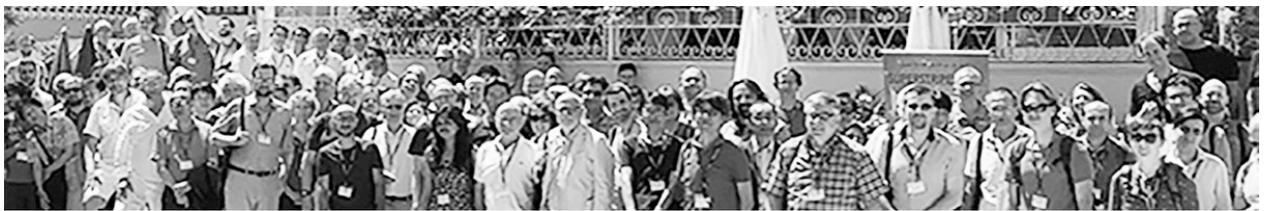

Fig. 4. Int. conference *Superstripes 2019*, Ischia, Italy June 23-29, 2019

This is a typical scenario near a Lifshitz transition for the appearing of a new Fermi surface spot described for organics [86] and predicted also for pressurized sulfur hydrides [93-95]. Advances on quantum mechanics of orbital physics which is today a hot topic have been obtained by Oleś group showing orbital dilution in $d^4$ oxides by tuning crystal field potential [96] and magnetic properties of crystalline and ferroelectric layered rare-earth-titanate by Kuznetsov et al. [97]. Strain Induced orbital dynamics across the metal insulator transition in thin $VO_2$ films [98] provide further information on nanoscale phase separation in the correlation-driven insulator-metal transition in vanadium dioxide [99-102]. The theoretical studies on nanoscale phase separation observed in cuprates [103-106] have focused on the coexistence of a metallic phase with a spin density wave phase controlled by pressure [107] of the nucleation of magnetic micro-inhomogeneity induced by an electric field [108] the order-disorder transition in a system made of two electronic components induced by nearest-neighbor repulsion [109].

In the rapidly developing field of quantum electronics, key results have been obtained on the control of quantum fluctuations in nanowires and on voltage fluctuations in coupled superconducting nanowires by Zaikin group [110-111] and on kinetic Inductance in superconducting microstructures by Shein et al [112]. Relevant memristive properties of perovskite superconductors have been found by Tulina, and Ivanov [113]. The theoretical work of Tanatar group has focused his interest on exchange-correlation effects in 2D dipolar Fermi liquid [114],

Nesselrodt et al [115] studied the Falicov-Kimball model for two components components, the Pepin group [116] studied the fractionalization of the pair density waves and the pseudo gap while the Seibold group reported new studies on the time-dependent Gutzwiller approximation [117].

## 5. Conclusions

The large number of participants to Superstripes 2019 conference (Fig. 4) has clearly shown the maturity of the novel paradigm "Superconductivity in Quantum Complex Matter" (SQCM) for the physics high temperature superconductivity based on fine tuning of nanoscale heterogeneity of complex matter which today is becoming the paradigm for the majority of scientists in this field. Now multi-band Hubbard models are currently used for cuprates, organics and iron based superconductors. Multiorbital models are currently used for diborides, doped fullerene and pressurized hydrides. It is now clear for that the inclusion of lattice instabilities of perovskites [118] and the anisotropic strain [119] is needed to understand the anisotropic multi gap superconductivity in strongly correlated systems. Moreover, today there is a high interest on electron-phonon interaction [43,95] and on lattice heterogeneity as proposed by Alex [120]. A new rapidly developing field is at the crossing point between the research a) on mixed boson-fermion systems in ultracold gases [121] b) shape resonances in multigap superconductivity near Lifshitz transitions in complex heterostructures and c) percolation of filamentary superconductivity in a granular landscape showing mesoscale correlated disorder [105-106] after the accumulated information on complex spatial distribution of defects, strain fluctuations, SDW puddles [122] after many works made in these last ten years [123-127].

## References

1. Bednorz, J. G., & Müller, K. A. (1986). Possible highTc superconductivity in the Ba− La− Cu− O system. *Zeitschrift für Physik B Condensed Matter*, *64*(2), 189-193.
2. Bednorz, J. G., Müller, K. A. (1988). Perovskite-type oxides—the new approach to high-$T_c$ superconductivity. *Reviews of Modern Physics*, *60*(3), 585.
3. Hock, K. H., Nickisch, H., Thomas, H. (1983). Jahn-Teller Effect in Itinerant Electron-Systems-the Jahn-Teller Polaron. *Helvetica Physica Acta*, *56*(1-3), 237-243
4. Chakraverty, B. K. (1981). Bipolarons and superconductivity. *Journal de Physique*, *42*(9), 1351-1356.
5. Bussmann-Holder, A., Bishop, A. R., & Egami, T. (2005). Relaxor ferroelectrics and intrinsic inhomogeneity. *EPL (Europhysics Letters)*, *71*(2), 249.
6. Shevchenko, S. N. (2019). Mesoscopic Physics meets Quantum Engineering. World Scientific Publishing Company Pte. Limited. ISBN: 978-981-12-0139-4
7. Zagoskin, A. M. (2011). Quantum engineering: theory and design of quantum coherent structures. Cambridge University Press
8. Kivelson S.A., Emery V.J. (2002) Stripe Liquid, Crystal, and Glass Phases of Doped Antiferromagnets. In: Bianconi A., Saini N.L. (eds) Stripes and Related Phenomena. Selected Topics in Superconductivity, vol 8. Springer, Boston, MA

9. Salamon, M. B., Bardeen, J. (1987). Comment on" Bulk Superconductivity at 91 K in Single-Phase Oxygen-Deficient Perovskite Ba$_2$YCu$_3$O$_{9-\delta}$" *Physical Review Letters*, *59*(22), 2615
10. Friedberg, R., & Lee, T. D. (1989). Boson-fermion model of superconductivity. *Physics Letters A*, *138*(8), 423-427
11. Heeger, A. J., & Yu, G. (1993). High-T c superconductors: Disordered metals with pairing via polarizability from localized states near the mobility edge. *Physical Review B*, *48*(9), 6492.
12. Ginzburg, V. L. (1991). High-temperature superconductivity (history and general review). *Soviet Physics Uspekhi*, *34*(4), 283
13. Ginzburg, V. L. (1992). Once again about high-temperature superconductivity. *Contemporary Physics*, *33*(1), 15-23
14. Ginzburg V.L. (1996) Bill Little and High Temperature Superconductivity. In: Cabrera B., Gutfreund H., Kresin V. (eds) From High-Temperature Superconductivity to Microminiature Refrigeration. Springer, Boston, MA
15. Ginzburg, V. L. (2004). Nobel Lecture: On superconductivity and superfluidity (what I have and have not managed to do) as well as on the "physical minimum" at the beginning of the XXI century. *Reviews of Modern Physics*, *76*(3), 981
16. Goodenough, J. B., & Zhou, J. S. (1997). Vibronic states in La$_{2-x}$Ba$_x$CuO$_4$. *Journal of Superconductivity*, *10*(4), 309-314.
17. Goodenough, J. B., & Zhou, J. S. (1997). New forms of phase segregation. *Nature*, *386*(6622), 229-230.
18. Goodenough, J.B. (2002). Ordering of bond length fluctuations in the copper-oxide superconductors. *EPL (Europhysics Letters)*, *57*(4), 550.
19. Bersuker, G. I., & Goodenough, J. B. (1997). Large low-symmetry polarons of the high-Tc copper oxides: formation, mobility and ordering. *Physica C: Superconductivity*, *274*(3-4), 267-285.
20. Deutscher, G., & de Gennes, P. G. (2007). A spatial interpretion of emerging superconductivity in lightly doped cuprates. *Comptes Rendus Physique*, *8*(7-8), 937-941
21. Barišić, S., Batistić, I., & Friedel, J. (1987). Electron-Phonon Model for High-Tc Layered-Metal Oxides. *EPL (Europhysics Letters)*, *3*(11), 1231.
22. Kresin V. Z. and Friedel J. (2011) Dynamic coexistence of various configurations: Clusters vs. nuclei *EPL (Europhysics Letters)*, **93** 13002 doi:10.1209/0295-5075/93/13002
23. Deutscher, G. (2012). The role of Cu-O bond length fluctuations in the high temperature superconductivity mechanism. *Journal of Applied Physics*, *111*(11), 112603.
24. Jorgensen, A. J., Dabrowski, B., Pei, S., Hinks, D. G., Soderholm, L., Morosin, B., ... & Ginley, D. S. (1988). Superconducting phase of La$_2$CuO$_{4+\delta}$: A superconducting composition resulting from phase separation. *Physical Review B*, *38*(16), 11337
25. Bishop, A. R., Martin, R. L., Müller, K. A., & Tešanović, Z. (1989). Superconductivity in oxides: Toward a unified picture. *Zeitschrift für Physik B Condensed Matter*, *76*(1), 17-24.
26. Teller E. (1989) Adaptation of the Theory of Superconductivity to the Behavior of Oxides. in Greiner W., Stöcker H. (eds) The Nuclear Equation of State. NATO ASI Series (Series B: Physics), vol 216a. Springer, Boston, doi:10.1007/978-1-4613-0583-5_37
27. Ashkenazi J., Vacaru D., and Kuper C. G. (1991) "Search of the correct microscopic theory for high temperature cuprate superconductors" in High Temperature Superconductivity edited by J. Ashkenazi et al. Plenum Press New York 1991, p. 569-582
28. Radaelli, P. G., Jorgensen, J. D., Kleb, R., Hunter, B. A., Chou, F. C., & Johnston, D. C. (1994). Miscibility gap in electrochemically oxygenated La 2 CuO 4+ δ. *Physical Review B*, *49*(9), 6239
29. Hammel, P. C., & Scalapino, D. J. (1996). Local microstructure and the cuprate spin gap puzzle. *Philosophical Magazine B*, *74*(5), 523-528.
30. Egami, T. (1996). Electron-lattice interaction in cuprates. *Journal of Low Temperature Physics*, *105*(3-4), 791-800.
31. Egami, T. (2001). Inhomogeneous charge state in HTSC cuprates and CMR manganites. *Physica C: Superconductivity and its applications*, *364*, 441-445.
32. Egami, T., & Louca, D. (2002). Charge localization in CMR manganites: Renormalization of polaron energy by stress field. *Physical Review B*, *65*(9), 094422
33. Egami, T., McQueeney, R. J., Petrov, Y., Shirane, G., & Endoh, Y. (2002). Low-Temperature Phonon Anomalies in Cuprates. In *Stripes and Related Phenomena* (pp. 191-197). Springer, Boston, MA.
34. Phillips, J. C., & Thorpe, M. F. (Eds.). (2006). *Phase transitions and self-organization in electronic and molecular networks*. Springer Science & Business Media.
35. Wakimoto, S., Kimura, H., Fujita, M., Yamada, K., Noda, Y., Shirane, G., ... & Birgeneau, R. J. (2006). Incommensurate lattice distortion in the high temperature tetragonal phase of La$_{2-x}$(Sr, Ba)$_x$CuO$_4$. *Journal of the Physical Society of Japan*, *75*(7), 074714-074714.
36. Haase, J., Slichter, C. P., Stern, R., Milling, C. T., & Hinks, D. G. (2000). NMR evidence for spatial modulations in the cuprates. *Journal of Superconductivity*, *13*(5), 723-726

37. Klein, N., Tellmann, N., Schulz, H., Urban, K., Wolf, S. A., & Kresin, V. Z. (1993). Evidence of two-gap s-wave superconductivity in YBa 2 Cu 3 O 7− x from microwave surface impedance measurements. *Physical Review Letters*, *71*(20), 3355.
38. Kresin, V. Z., & Wolf, S. A. (2009). Colloquium: electron-lattice interaction and its impact on high Tc superconductivity. *Reviews of Modern Physics*, *81*(2), 481.
   two component
39. Kresin, V. Z., Ovchinnikov, Y. N., & Wolf, S. A. (2006). Inhomogeneous superconductivity and the "pseudogap" state of novel superconductors. *Physics Reports*, *431*(5), 231-259.
40. Ashcroft, N. W. (1992). Exotic Atoms in Condensed Matter: Conclusions. In *Exotic Atoms in Condensed Matter* (pp. 297-305). Springer, Berlin, Heidelberg.
41. Ashcroft, N. W. (2006). Symmetry and higher superconductivity in the lower elements. In *Symmetry and Heterogeneity in High Temperature Superconductors* (pp. 3-20). Springer, Dordrecht
42. Gor'kov, L. P. (2000). Phase separation in a two-component model for cuprates. *Journal of Superconductivity*, *13*(5), 765-769.
43. Gor'kov, L. P., & Teitel'Baum, G. B. (2015). Two-component energy spectrum of cuprates in the pseudogap phase and its evolution with temperature and at charge ordering. *Scientific reports*, *5*, 8524
44. Iachello, F. (2006). Symmetry of High-$T_c$ Superconductors. In *Symmetry and Heterogeneity in High Temperature Superconductors* (pp. 165-180). Springer, Dordrecht.
45. Müller, K. A. (1991) The first five years of high-Tc superconductivity. *Physica C: Superconductivity*, *185*, 3-10 https://doi.org/10.1016/0921-4534(91)91942-W
46. Muller, K. A., & Benedek, G. (Eds.). (1993). *Phase separation in cuprate superconductors*. Singapore: World Scientific.
47. Sigmund, E., & Müller, K. A. (Eds.). (2012). *Phase Separation in Cuprate Superconductors: Proceedings of the second international workshop on "Phase Separation in Cuprate Superconductors" September 4–10, 1993, Cottbus, Germany*. Springer Science & Business Media.
48. Bianconi, A., & Missori, M. (1994). The instability of a 2D electron gas near the critical density for a Wigner polaron crystal giving the quantum state of cuprate superconductors. *Solid State Communications*, *91*(4), 287-293
49. Bianconi, A., Missori, M., Oyanagi, H., Yamaguchi, H., Ha, D. H., Nishiara, Y., & Della Longa, S. (1995). The measurement of the polaron size in the metallic phase of cuprate superconductors. *EPL (Europhysics Letters)*, *31*(7), 411
50. Bianconi, A. (1998). European Patent N. 0733271" High $T_c$ superconductors made by metal heterostuctures at the atomic limit" (priority date 7 Dec 1993), published in European Patent Bulletin 98/22, May 27 1998)
51. Bianconi, A. (2001). *U.S. Patent No. 6,265,019*. Washington, DC: U.S. Patent and Trademark Office.
52. Lanzara, A., Saini, N. L., Rossetti, T., Bianconi, A., Oyanagi, H., Yamaguchi, H., & Maeno, Y. (1996). Temperature dependent local structure of the $CuO_2$ plane in the 1/8 doped $La_{1.875}Ba_{0.125}CuO_4$ system. *Solid State Communications*, *97*(2), 93-96.
53. Bianconi, A., Saini, N. L., Rossetti, T., Lanzara, A., Perali, A., Missori, M., ... & Ha, D. H. (1996). Stripe structure in the $CuO_2$ plane of perovskite superconductors. *Physical Review B*, *54*(17), 12018.
54. Bianconi, A., Saini, N.L., Lanzara, A., Missori, M., Rossetti, T., Oyanagi, H., ... & Ito, T. (1996). Determination of the Local Lattice Distortions in the $CuO_2$-Plane of $La_{1.85}Sr_{0.15}CuO_4$. *Physical Review Letters*, *76*(18), 3412.
55. Bianconi, A., Lusignoli, M., Saini, N. L., Bordet, P., Kvick, Å., & Radaelli, P. G. (1996). Stripe structure of the $CuO_2$ plane in $Bi_2Sr_2CaCu_2O_{8+\delta}$ by anomalous X-ray diffraction. *Physical Review B*, *54*(6), 4310.
56. Perali, A., Bianconi, A., Lanzara, A., & Saini, N. L. (1996). The gap amplification at a shape resonance in a superlattice of quantum stripes: A mechanism for high $T_c$. *Solid State Communications*, *100*(3), 181-186.
57. Bianconi, A., Valletta, A., Perali, A., & Saini, N. L. (1997). High Tc superconductivity in a superlattice of quantum stripes. *Solid State Communications*, *102*(5), 369-374.
58. Valletta, A., Bianconi, A., Perali, A., & Saini, N. L. (1997). Electronic and superconducting properties of a superlattice of quantum stripes at the atomic limit. *Zeitschrift für Physik B Condensed Matter*, *104*(4), 707-713.
59. Bianconi, A., & Saini, N. L. (Eds.). (2001). *Stripes and related phenomena*. Springer Science & Business Media ISBN**:** 978-0306464195
60. Müller K.A. (2002) From Phase Separation to Stripes. In: Bianconi A., Saini N.L. (eds) Stripes and Related Phenomena. Selected Topics in Superconductivity, vol 8. Springer, Boston, MA https://doi.org/10.1007/0-306-47100-0_1
61. Bianconi, A., Valletta, A., Perali, A., & Saini, N. L. (1998). Superconductivity of a striped phase at the atomic limit. *Physica C: Superconductivity*, *296* (3-4), 269-280
62. Saini, N. L., Avila, J., Bianconi, A., Lanzara, A., Asensio, M. C., Tajima, S., ... & Koshizuka, N. (1997). Topology of the pseudogap and shadow bands in $Bi_2Sr_2CaCu_2O_{8+\delta}$ at optimum doping. *Physical Review Letters*, *79*(18), 3467.
63. Bianconi A., Agrestini S., Bianconi G., Di Castro D., Saini N.L. (2002) Lattice-Charge Stripes in the High-$T_c$ Superconductors. In: Bianconi A., Saini N.L. (eds) Stripes and Related Phenomena. Selected Topics in Superconductivity, vol 8. Springer, Boston, MA

64. Oyanagi H., Bianconi A. eds, Physics in Local Lattice Distortions *AIP Conference* (Proc. of Int. Conf. LLD2K Ibaraki (Japan) 23-26 July 2000), New York, United States *AIP Conference Proceedings* vol. **554** (2001) isbn: 9781563969843
https://www.bookdepository.com/Physics-Local-Lattice-Distortions-Hiroyuki-Oyanagi/9781563969843
65. Bianconi, A., Di Castro, D., Bianconi, G., & Saini, N. L., The strain quantum critical point for superstripes, *AIP Conference Proceedings* **554**, 124 (2001); https://doi.org/10.1063/1.1363067
66. Di Castro, D., Bianconi, G., Colapietro, M., Pifferi, A., Saini, N. L., Agrestini, S., & Bianconi, A. (2000). Evidence for the strain critical point in high $T_c$ superconductors. *Eur. Phys. J. B* **18,** 617–624 (2000) https://doi.org/10.1007/s100510070010 (Received 25 September 2000 published Dec 2000
67. Bianconi, A., Saini, N. L., Agrestini, S., Castro, D. D., & Bianconi, G. (2000). The strain quantum critical point for superstripes in the phase diagram of all cuprate perovskites. *International Journal of Modern Physics B*, *14*(29n31), 3342-3355. https://doi.org/10.1142/S0217979200003812
68. Bianconi A., Superstripes. *International Journal of Modern Physics B*, 14 (29n31), 3289-3297. (2000). (published 1 March 200) https://doi.org/10.1142/S0217979200003769
69. Bianconi, A., Bianconi, G., Caprara, S., Di Castro, D., Oyanagi, H., & Saini, N. L. (2000). The stripe critical point for cuprates. *Journal of Physics: Condensed Matter*, *12*(50), 10655. https://doi.org/10.1088/0953-8984/12/50/326 (Received 21 Sept 2000, published Dec 2000)
70. Nagamatsu, J., Nakagawa, N., Muranaka, T., Zenitani, Y., & Akimitsu, J. (2001). Superconductivity at 39 K in magnesium diboride. *Nature*, *410*(6824) 63–64 (2001). (received 24 Jan 2001-published on Mar 1, 2001) https://doi.org/10.1038/35065039
71. Agrestini, S., Di Castro, D., Sansone, M., Saini, N. L., Saccone, A., De Negri, S., Giovannini, M., Colapietro, M., Bianconi, A. (2001). High $T_c$ superconductivity in a critical range of micro-strain and charge density in diborides. *J. Phys.: Condens. Matter* **13** 11689 doi: 10.1088/0953-8984/13/50/328. (Received 1 Mar 2001; Published Nov 2001)
72. Bianconi, A., Di Castro, D., Agrestini, S., Campi, G., Saini, N. L., Saccone, A., ... & Giovannini, M. (2001). A superconductor made by a metal heterostructure at the atomic limit tuned at the shape resonance': $MgB_2$. *Journal of Physics: Condensed Matter*, *13*(33), 7383.
73. Bauer, E., Paul, C., Berger, S., et al. (2001). Thermal conductivity of superconducting $MgB_2$. *Journal of Physics: Condensed Matter*, *13*(22), L487. https://doi.org/10.1088/0953-8984/13/22/107 (Received 25 Apr 2001 published Jun 2001)
74. Agrestini, S., Metallo, C., Filippi, M., Simonelli, L., Campi, G., Sanipoli, C., ... & Latini, A. (2004). Substitution of Sc for Mg in $MgB_2$: Effects on transition temperature and Kohn anomaly. *Physical Review B*, *70*(13), 134514
75. Simonelli, L., Palmisano, V., Fratini, M., Filippi, M., Parisiades, P., Lampakis, D., ... & Bianconi, A. (2009). Isotope effect on the E2g phonon and mesoscopic phase separation near the electronic topological transition in Mg1− xAlxB2. *Physical Review B*, *80*(1), 014520.
76. Alarco, J. A., Almutairi, A., & Mackinnon, I. D. (2019). Progress Towards a Universal Approach for Prediction of the Superconducting Transition Temperature. *Journal of Superconductivity and Novel Magnetism*, doi:10.1007/s10948-019-05384-2
77. Campi, G., Cappelluti, E., Proffen, T., et al. (2006). Study of temperature dependent atomic correlations in $MgB_2$. *The European Physical Journal B-Condensed Matter and Complex Systems*, *52*(1), 15-21
78. Palmisano, V., Simonelli, L., Puri, A., et al. (2008). Controlling mesoscopic phase separation near electronic topological transitions via quenched disorder in ternary diborides. *Journal of Physics: Condensed Matter*, *20*(43), 434222
79. Kamihara, Y., Watanabe, T., Hirano, M., & Hosono, H. (2008). Iron-based layered superconductor La [O1-x F x] FeAs (x= 0.05− 0.12) with T c= 26 K. *Journal of the American Chemical Society*, *130*(11), 3296-3297.
80. Dagotto, E., Moreo, A., Nicholson, A., Luo, Q., Liang, S., & Zhang, X. (2011). Properties of the multiorbital Hubbard models for the iron-based superconductors. *Frontiers of Physics*, *6*(4), 379-397.
81. Innocenti, D., Poccia, N., Ricci, A., et al. (2010). Resonant and crossover phenomena in a multiband superconductor: Tuning the chemical potential near a band edge. *Physical Review B*, *82*(18), 184528.
82. Innocenti, D., Caprara, S., Poccia, N., Ricci, A., Valletta, A., & Bianconi, A. (2010). Shape resonance for the anisotropic superconducting gaps near a Lifshitz transition: the effect of electron hopping between layers. *Superconductor Science and Technology*, *24*(1), 015012.
83. Torsello, D., Ummarino, G.A., Gerbaldo, R. *et al.* Eliashberg Analysis of the Electrodynamic Response of Ba(Fe$_{1-x}$Rh$_x$)$_2$As$_2$ Across the s$_\pm$ to s$_{++}$ Order Parameter Transition. *J Supercond Nov Magn* (2019) doi:10.1007/s10948-019-05368-2
84. Pal, A., Chinotti, M., Chu, J. H., Kuo, H. H., Fisher, I. R., Degiorgi, L. (2019). Anisotropic Superconducting Gap in Optimally Doped Iron–Based Material. *Journal of Superconductivity and Novel Magnetism*, doi:10.1007/s10948-019-05390-4
85. Ptok, A., Kapcia, K.J., Sternik, M. *et al.* Superconductivity of KFe$_2$As$_2$ Under Pressure: Ab Initio Study of Tetragonal and Collapsed Tetragonal Phases. *J Supercond Nov Magn* (2020). doi:10.1007/s10948-020-05454-w

86. Mazziotti, M. V., Valletta, A., Campi, G., Innocenti, D., Perali, A., & Bianconi, A. (2017). Possible Fano resonance for high-Tc multi-gap superconductivity in p-Terphenyl doped by K at the Lifshitz transition. *EPL (Europhysics Letters)*, *118*(3), 37003.
87. Tsuchiya, S., Mertelj, T., Mihailovic, D. *et al.* Ultrafast Carrier Dynamics in an Organic Superconductor κ-(BEDT-TTF)$_2$Cu[N(CN)$_2$]Br by Spectrally Resolved Pump-Probe Spectroscopy. *J Supercond Nov Magn* (2019) doi:10.1007/s10948-019-05382-4
88. Nakagawa, K., Tsuchiya, S., Taniguchi, H. *et al.* Polarized Time-Resolved Spectroscopy of Electronic Phase Separation in a Dimer-Mott Organic Insulator. *J Supercond Nov Magn* (2020). https://doi.org/10.1007/s10948-019-05385-1
89. Bianconi, A., Poccia, N., Sboychakov, A. O., Rakhmanov, A. L., & Kugel, K. I. (2015). Intrinsic arrested nanoscale phase separation near a topological Lifshitz transition in strongly correlated two-band metals. *Superconductor Science and Technology*, *28*(2), 024005
90. Agrestini, S., Saini, N. L., Bianconi, G., & Bianconi, A. (2003). The strain of CuO$_2$ lattice: the second variable for the phase diagram of cuprate perovskites. *Journal of Physics A: Mathematical and General*, *36*(35), 9133.
91. Caprara S. (2019) The Ancient Romans' Route to Charge Density Waves in Cuprates *Condens. Matter* 2019, *4*(2), 60; https://doi.org/10.3390/condmat4020060
92. Aoki, H. Theoretical Possibilities for Flat Band Superconductivity. *J Supercond Nov Magn* (2020). doi:10.1007/s10948-020-05474-6
93. Bianconi, A., & Jarlborg, T. (2015). Superconductivity above the lowest Earth temperature in pressurized sulfur hydride. *EPL (Europhysics Letters)*, *112*(3), 37001.
94. Jarlborg, T., & Bianconi, A. (2016). Breakdown of the Migdal approximation at Lifshitz transitions with giant zero-point motion in the H$_3$S superconductor. *Scientific reports*, *6*, 24816.
95. Gor'kov, L. P., & Kresin, V. Z. (2016). Pressure and high-Tc superconductivity in sulfur hydrides. *Scientific reports*, *6*(1), 1-7.
96. Brzezicki, W., Forte, F., Noce, C., Cuoco, M., Oleś, A. M. Tuning Crystal Field Potential by Orbital Dilution in d$^4$ Oxides. J Supercond Nov Magn (2019). doi:10.1007/s10948-019-05386-0
97. Kuznetsov, A.V., Churkin, O.A., Popov, V.V. *et al.* Magnetization of Crystalline and Amorphous Phases of R$_2$Ti$_2$O$_7$ and R$_2$Zr$_2$O$_7$ (R = Gd, Dy, Tb). *J Supercond Nov Magn* (2020). doi:10.1007/s10948-019-05388-y
98. D'Elia, A., Rezvani, S., Cossaro, A. *et al.* Strain induced orbital dynamics across the metal insulator transition in thin VO$_2$/TiO$_2$ (001) films. *J Supercond Nov Magn* (2020). https://doi.org/10.1007/s10948-019-05378-0
99. Gray, A. X., Jeong, J., Aetukuri, N. P., Granitzka, P., Chen, Z., Kukreja, R., ... & Marcus, M. A. (2016). Correlation-driven insulator-metal transition in near-ideal vanadium dioxide films. *Physical Review Letters*, *116*(11), 116403.
100. Bianconi, A. (1982). Multiplet splitting of final-state configurations in x-ray-absorption spectrum of metal VO$_2$: Effect of core-hole-screening, electron correlation, and metal-insulator transition. *Physical Review B*, *26*(6), 2741.
101. Gioacchino, D. D., Marcelli, A., Puri, A., Zou, C., Fan, L., Zeitler, U., & Bianconi, A. (2017). Metastability phenomena in VO$_2$ thin films. *Condensed Matter*, *2*(1), 10.
102. Marcelli, A., Coreno, M., Stredansky, M., Xu, W., Zou, C., Fan, L., ... & Bianconi, A. (2017). Nanoscale phase separation and lattice complexity in VO$_2$: The metal–insulator transition investigated by XANES via Auger electron yield at the vanadium L$_{2,3}$-edge and resonant photoemission. *Condensed Matter*, *2*(4), 38.
103. Campi, G., Bianconi, A., Poccia, N., Bianconi, G., Barba, L., Arrighetti, G., ... & Burghammer, M. (2015). Inhomogeneity of charge-density-wave order and quenched disorder in a high-T c superconductor. *Nature*, *525*(7569), 359-362.
104. Ricci, A., Poccia, N., Joseph, B., Innocenti, D., Campi, G., Zozulya, A., ... & Takeya, H. (2015). Direct observation of nanoscale interface phase in the superconducting chalcogenide K$_x$Fe$_{2-y}$Se$_2$ with intrinsic phase separation. *Physical Review B*, *91*(2), 020503.
105. Campi, G., Innocenti, D., & Bianconi, A. (2015). CDW and similarity of the Mott insulator-to-metal transition in cuprates with the gas-to-liquid-liquid transition in supercooled water. *Journal of Superconductivity and Novel Magnetism*, *28*(4), 1355-1363.
106. Campi, G., & Bianconi, A. (2016). High-Temperature superconductivity in a hyperbolic geometry of complex matter from nanoscale to mesoscopic scale. *Journal of Superconductivity and Novel Magnetism*, *29*(3), 627-631.
107. Rakhmanov, A.L., Kugel, K.I. & Sboychakov, A.O. Coexistence of spin density wave and metallic phases under pressure. *J Supercond Nov Magn* (2020). doi:10.1007/s10948-019-05379-z
108. Kulikova, D.P., Nikolaeva, E.P., Ren, W. *et al.* Electric field–induced nucleation of magnetic micro-inhomogeneities and bubble domain lattices. *J Supercond Nov Magn* (2020). doi:10.1007/s10948-019-05370-8
109. Kapcia, K. J., & Majewska-Albrzykowska, K. (2020). Order-disorder transition in the half-filled two-component lattice fermion model with nearest-neighbor repulsion. *Journal of Superconductivity and Novel Magnetism*, 1-8. https://doi.org/10.1007/s10948-020-05453-x
110. Radkevich, A., Semenov, A.G. & Zaikin, A.D. Topology-Controlled Phase Coherence and Quantum Fluctuations in Superconducting Nanowires. *J Supercond Nov Magn* (2020). https://doi.org/10.1007/s10948-019-05381-5

111. Latyshev, A., Semenov, A.G. & Zaikin, A.D. Voltage Fluctuations in a System of Capacitively Coupled Superconducting Nanowires. *J Supercond Nov Magn* (2020). https://doi.org/10.1007/s10948-019-05402-3
112. Shein, K.V., Emelyanova, V.O., Logunova, M.A. *et al.* Kinetic inductance in superconducting microstructures. *J Supercond Nov Magn* (2020). https://doi.org/10.1007/s10948-019-05401-4
113. Tulina, N.A., Ivanov, A.A. Memristive properties of oxide-based high-temperature superconductors. *J Supercond Nov Magn* (2020). https://doi.org/10.1007/s10948-019-05383-3
114. Seydi, I., Abedinpour, S. H., Asgari, R., Tanatar, B. (2019). Exchange-correlation effects and the quasiparticle properties in a two-dimensional dipolar fermi liquid. *Journal of Superconductivity and Novel Magnetism*, 1-6. doi:10.1007/s10948-019-05371-7
115. Nesselrodt, R. D., Canfield, J., & Freericks, J. K. (2020). Comparison between the f-electron and conduction-electron density of states in the falicov-kimball model at low temperature. *J Supercond Nov Magn* (2020). https://doi.org/10.1007/s10948-019-05400-5
116. Grandadam, M., Chakraborty, D. & Pépin, C. Fractionalizing a local pair density wave: a good "recipe" for opening a pseudo-gap. *J Supercond Nov Magn* (2020). doi:10.1007/s10948-019-05380-6
117. Noatschk, K., Martens, C., & Seibold, G. (2020). Time-dependent Gutzwiller approximation: theory and applications. *Journal of Superconductivity and Novel Magnetism*, doi.: 10.1007/s10948-019-05406-z
118. Makarov, I. A., Gavrichkov, V. A., Shneyder, E. I., Nekrasov, I. A., Slobodchikov, A. A., Ovchinnikov, S. G., & Bianconi, A. (2019). Effect of $CuO_2$ Lattice Strain on the Electronic Structure and Properties of High-$T_c$ Cuprate Family. *Journal of Superconductivity and Novel Magnetism*, *32*(7), 1927-1935.
119. Vladimir A. Gavrichkov, Yury Shan'ko, Natalia G. Zamkova, and Antonio Bianconi, Is there any hidden symmetry in the stripe structure of perovskite high-temperature superconductors? *The Journal of Physical Chemistry Letters* **2019** *10* (8), 1840-1844 DOI: 10.1021/acs.jpclett.9b00513
120. Egami, T. (2017) Alex and the Origin of High-Temperature Superconductivity. In *High-Tc Copper Oxide Superconductors and Related Novel Materials* (pp. 35-46). Springer, Cham
121. Kagan, M. Y., & Bianconi, A. (2019). Fermi-Bose mixtures and BCS-BEC crossover in high-Tc superconductors. *Condensed Matter*, *4*(2), 51.
122. Campi, G., Poccia, N., Joseph, B., Bianconi, A., Mishra, S., Lee, J., ... & Trabant, C. (2019). Direct visualization of spatial inhomogeneity of spin stripes order in $La_{1.72}Sr_{0.28}NiO$. *Condensed Matter*, *4*(3), 77.
123. Zaanen, J. (2010). The benefit of fractal dirt. *Nature*, *466*(7308), 825-826.
124. Littlewood, P. (2011). An X-ray oxygen regulator. *Nature materials*, *10*(10), 726-727
125. Poccia, N., Ricci, A., Campi, G., Fratini, M., Puri, A., Di Gioacchino, D., ... & Aeppli, G. (2012). Optimum inhomogeneity of local lattice distortions in $La_2CuO_{4+y}$. *Proceedings of the National Academy of Sciences*, *109*(39), 15685-15690
126. Dagotto, E. (2013). Colloquium: The unexpected properties of alkali metal iron selenide superconductors. *Reviews of Modern Physics*, *85*(2), 849.
127. Phillips, J. C. (2014). Ineluctable complexity of high temperature superconductivity elucidated. *Journal of Superconductivity and Novel Magnetism*, *27*(2), 345-347.